# Immigrant Women and the COVID-19 Pandemic: An Intersectional Analysis of Frontline Occupational Crowding

*Sarah F Small, University of Utah[1]*
*Yana van der Meulen Rodgers, Rutgers University*
*Teresa Perry, California State University San Bernardino*

ABSTRACT: This paper examines changes in occupational crowding of immigrant women in frontline industries in the United States during the onset of COVID-19, and we contextualize their experiences against the backdrop of broader race-based and gender-based occupational crowding. Building on the occupational crowding hypothesis, which suggests that marginalized workers are crowded in a small number of occupations to prop up wages of socially-privileged workers, we hypothesize that immigrant, Black, and Hispanic workers were shunted into frontline work to prop up the health of others during the pandemic. Our analysis of American Community Survey microdata indicates that immigrant workers, particularly immigrant women, were increasingly crowded in frontline work during the onset of the pandemic. We also find that US-born Black and Hispanic workers disproportionately faced COVID-19 exposure in their work, but were not increasingly crowded into frontline occupations following the onset of the pandemic. The paper also provides a rationale for considering the occupational crowding hypothesis along the dimensions of both wages and occupational health.

[1] Corresponding author: Sarah Small, 260 Central Campus Drive, Gardner Commons, Salt Lake City, Utah 84112, sarah.small@utah.edu. This research was funded in part by the Black Bodies Black Health Research Grant from the Institute for the Study of Global Racial Justice at Rutgers University. We thank Debra Lancaster, Michelle Stephens, and Enobong (Anna) Branch for their helpful comments and suggestions.

# I. INTRODUCTION

Occupational segregation and discrimination have long marred the United States labor market and have contributed to wage and health gaps along axes of gender, race, ethnicity, and nativity. Occupational segregation became especially problematic during the COVID-19 pandemic, where frontline essential workers were exposed to the virus at disproportionate rates and faced lower than average wages during the corresponding economic crisis. This paper seeks to understand changes in occupational crowding in frontline industries during the onset of COVID-19, and in particular whether occupational segregation shifted during the pandemic to push more vulnerable groups into more dangerous work.

The analysis uses American Community Survey microdata and an occupational crowding index measure that estimates the over- and under-representation of race and gender groups in occupations, while controlling for educational attainment. Results show that immigrant workers, particularly immigrant women, were increasingly crowded into frontline work during the onset of the pandemic. Our results highlight the increased exploitation of immigrant workers during COVID-19. We also find that US-born Black and Hispanic workers disproportionately faced COVID-19 exposure in their work.

We contend that occupational crowding in frontline industries is in line with theory from stratification economics, which views race- and ethnicity-based discrimination as a rational attempt on behalf of privileged groups to preserve their relative status and the material benefits which that status confers (Darity 2005, Seguino 2021). The ability for White, US-born workers to disproportionately abstain from frontline work indicates their inherent advantage in preserving their health in times of health crises. Building on the occupational crowding hypothesis, which suggests that marginalized workers are clustered in a small number of occupations to prop up wages of socially-privileged workers, we hypothesize that during the pandemic, marginalized immigrant workers were shunted into frontline work to prop up the health of others.

## II. CONCEPTUAL UNDERPINNINGS

A long line of studies in stratification and feminist economics has found that Black workers, immigrant workers with darker skin tones, and women tend to be crowded primarily in low-wage occupations (Dávila, Mora, and Stockly 2011, Holder 2017, Holder 2018, Wilson 2021), not just in the United States, but also in many other parts of the world (Seguino and Braunstein 2019). Similar research has found that immigrant workers in the United States are concentrated in low-wage occupations (Joassart-Marcelli 2014) and occupations with major health hazards (McCauley 2005; Flynn et al. 2015).

Ultimately, the occupational segregation of immigrants and members of underrepresented racial/ethnic groups and reliance on these groups to complete treacherous and undesirable work has an extensive history in the United States. The United States enslaved Black workers to complete arduous agricultural work (Jones, 2009), and it imported Chinese workers to construct the railroad systems (Boswell 1986).  More recently the U.S. has imported Hispanic and Latino workers for farm work that is associated with a long list of adverse health outcomes (Arcury and Martin 2009; Xiao et al. 2013; Stoecklin-Marois 2015; Pulgar et al. 2016). The United States continues to import low-income immigrant women to take on paid care work of children, the elderly, and disabled individuals, thus creating global care chains (Glenn 2000; Fraser 2017; Phillips-Cunningham 2019). Policies governing US immigration have enough latitude that US households can continue to employ an abundant supply of immigrant women as low-wage domestic workers (Chang 2000; Banks 2006). In the US, low-wage and time-intensive reproductive work is disproportionately performed by women of color and is enabled by legal, political, and economic exclusions that have prevented people of color from organizing and competing for better paying jobs (Hondagneu-Sotelo 2007). Undocumented women are frequently tracked into domestic work because of constraints they face from immigration policies combined with the lack of public regulation of private homes (Rodríguez 2007).

Immigrants are more likely to work in high-risk jobs with poor working conditions (Orrenius and Zavodny 2009). Davila, Mora, and Gonzalez (2011) explain that this is one contributing factor for the increase in on-the-job fatalities and injuries for Hispanic workers in the past two decades. Foreign born Hispanic workers have higher fatality rates than non-Hispanic and Hispanic workers who are native born (Richardson, Ruser, and Suarez 2003; Loh and Richardson 2004; Orrenius and Zavodny 2009). This pattern continued into the pandemic as employers in the United States relied heavily on immigrants to fill frontline essential work, and some sought to take advantage of immigrant workers' precarious status to further expand frontline essential work provisions.. For example, Liebert (2021) suggests using streamlined integration policies for immigrant workers so they can fill the ever-growing demand for healthcare workers. Other scholars have argued that legislation providing pathways to citizenship for essential workers demonstrates that immigrants were not considered to be worthy of security or belonging until they were proven 'essential' (Izu 2020).

Improving employment and citizenship opportunities for immigrants is essential for the well-being of migrants as well as the nation's economic success. However, the urgency to place immigrant workers in frontline work highlights employers' (and policymakers') preferences for recruiting exploitable migrant workers rather than improving wages and working conditions needed to attract and retain US-born workers. Indeed, previous research has found that employers seek to employ non-citizen workers because workers are often not free to leave sponsoring employers and are therefore more vulnerable to abuses (Anderson 2010).

Because the economic fallout from COVID-19 is not occupation-neutral, occupational segregation by race, nativity, and gender has become a major concern for health researchers and economists. Nationwide, scholars have shown that women, Black workers, Hispanic workers, and immigrant workers are overrepresented in frontline industries, meaning they were disproportionately exposed to COVID-19 (Hawkins 2020; Goldman et al., 2020; Blau et al. 2021). In fact, recent research has found that young immigrants faced exceptionally high COVID-19 mortality rates compared to young US-born citizens (Horner et al. 2022).

Recent research has also shown that frontline workers have higher odds of contracting COVID-19 compared to those in non-essential work (Do and Frank 2021). Healthcare workers in particular faced among the highest rates of COVID-19 exposure (Zhang 2021; Jin et al. 2021). In addition to an increased risk of COVID-19 exposure, frontline workers also experienced adverse mental health outcomes, including increases in stress, anxiety, and insomnia, as well as declines in overall well-being (Magnavita et al. 2021; Moitra et al. 2021; Newman et al. 2022). Because frontline essential workers tend to have low wages, if and when they contract the virus, they have fewer resources to fall back on.

Though made apparent by the pandemic, low wages in essential work is not a new phenomenon. Real wages in essential industries have declined relative to nonessential industries since 1983, and essential industries have consistently had lower levels of wage inequality than their nonessential counterparts (Walke 2021). Uneven rates of de-unionization can explain part of the decline in relative wages (Walke 2021). Workers in essential care service jobs – especially women – earn even less than other essential workers (Folbre et al. 2021).

Ultimately, the overrepresentation of immigrants, Black workers, and Hispanic workers in frontline occupations suggests that already vulnerable groups are being shunted into risky work while White workers and men are largely able to avoid such work. In what follows, we build on the theoretical contributions of stratification economics and the occupational crowding hypothesis. Stratification economics contends that there are material benefits to discrimination that accrue to those in positions of power and privilege (Chelwa et al. 2022). The occupational crowding hypotheses highlights wage dimensions of such material benefits, as marginalized workers are clustered in a small number of occupations to prop up wages of socially-privileged workers. We next suggest an extension of the occupational crowding framework in which marginalized workers are relegated to risky occupations to prop up the health of others.

III. THE OCCUPATIONAL CROWDING HYPOTHESIS: AN EXTENSION TO HEALTH ECONOMICS

The current version of the occupational crowding hypothesis is largely credited to Barbara Bergmann's work, which was met with mixed reception in economics because of its focus on group conflict and power as opposed to mainstream theories of discriminatory 'tastes and preferences' (Small 2022). In contrast to the mainstream, feminist and stratification economists have employed the occupational crowding hypothesis in research on racial disparities in labor markets (for instance, Holder 2018 and 2017; Hamilton and Darity 2012; Hamilton et al. 2011; Hamilton 2006; Gibson et al. 1998). In addition, the crowding hypothesis has also made its mark on studies on immigrant labor (Stevans 1998 and 1996).

Bergmann's occupational crowding hypothesis suggests the following: first, that Black workers are intentionally excluded from certain occupations. This exclusion limits the labor supply and therefore props up the wages of White workers for whom those occupations are reserved. As Black workers are crowded into a smaller number of occupations, the labor supply and competition for such jobs is relatively high, thus suppressing their wages (Bergmann 1971).

In line with Bergmann's crowding hypothesis, if Black or otherwise marginalized workers are crowded into low wage occupations in order to prop up White wages in other occupations, analogously, we suggest that marginalized bodies may be crowded into occupations with high health risks in order to prop up the ability for White bodies to minimize their own health risk exposure. We hypothesize that throughout the pandemic, White US-born consumers benefit from services provided by immigrant, Black, and Hispanic frontline workers, but face a substantially smaller risk of adverse health consequences in their own work. In the remainder of the paper, we use micro-data and an occupational crowding index to examine whether already marginalized workers were increasingly crowded into low-wage and high-risk jobs during the pandemic.

## IV. DATA AND METHODS

We first examine changes in the occupational crowding index among immigrant groups, then turn to US-born workers by race and ethnicity. We follow Holder (2017), Holder (2018), Hamilton et al. (2011), and Gibson et al. (1998), which adapted Bergmann's (1971) methodology, to measure occupational crowding. In our analysis of immigrant and US-born workers, the occupational crowding index is an occupation-specific ratio, where the numerator captures immigrant workers employed in the occupation j as a share of all workers in occupation j. The denominator captures immigrant workers with the educational attainment necessary for occupation j as a share of the working age population with the same education level.

$$\text{Crowding Index} = \frac{I_j / LF_j}{I_j^{edu} / CP_j^{edu}}$$

If the crowding index equals one, then immigrant workers are perfectly represented in the occupation. If the index is greater than one, then immigrant workers are overrepresented, or crowed, in the occupation. If it is less than one, then immigrant workers are underrepresented in that occupation.

Using American Community Survey (ACS) data from 2019 and 2020, we first calculate the occupational crowding index among immigrant workers in the United States for major occupational groups. We consider immigrants to be those who indicted they are either a naturalized citizen or a non-citizen, born outside the United States. Our measure of non-citizens includes undocumented immigrants, although there is no way to distinguish such immigrants in ACS data, and some have suggested undocumented immigrants may be undercounted in census data (Massey and Capoferro 2004).[2] In Appendix 1, we provide descriptive statistics on the birthplaces of immigrants in our sample.

The ACS data indicate that overall, the share of immigrants in the population and in the labor force changed little with the onset of the pandemic. In 2019, immigrants made up 11.6% of the survey's respondents and 15.9% of prime-age respondents (ages 25 to 64, inclusive) in the labor force. In 2020, these figures were 11.5% and 15.9%, respectively, suggesting that changes in occupational crowding indices are not driven by vast changes in the immigrant population or their labor force participation.

Our major occupation groups are calculated at the four-digit level of disaggregation consistent with the Standard Occupational Classification system. The years 2019 and 2020 are chosen in an attempt to capture differences in crowding indices from before and during the

---

pandemic. Because of reports that Black individuals and Hispanic individuals were undercounted in the 2020 ACS data (Wang 2022), we supplement our analysis with data from the Current Population Survey (CPS). Although the CPS has smaller annual samples, the survey allows us to capture more current information from 2021 and the early months of 2022. Results from the CPS calculations are presented in Appendix 2.

We calculate the crowding index following Holder's (2017) methodology and sample restrictions. Namely, we restrict the ages in our sample to 25-64 years of age (in order to avoid confounding factors in retirements and school enrollments (Hamilton et al., 2011)). In the numerator of the occupational crowding index, we include only employed individuals, while in the denominator (the portion attributable to educational attainment) we include individuals who are employed, unemployed, or out of the labor force. This step allows us to capture those who could have been working in each occupation given their educational attainment. In the denominator of the index, where an estimate for educational thresholds by occupation is necessary for the calculation, we calculate the $25^{th}$ and $90^{th}$ percentiles of educational attainment for all sample respondents in each occupation category. We then use this range to determine the number of immigrant workers who possess education within the threshold for the occupation. Determining the share of immigrant workers who possesses education levels for each occupation allows us to estimate the share of immigrant workers expected to be in that occupation.

This attempt to control for differences in educational attainment is common in the occupational segregation literature (Holder 2017). While this method is useful for comparing discrimination in occupational employment patterns, we caution that it is also controlling for discrimination occurring earlier in the life cycle. For instance, it has been well documented that Black girls are systematically discouraged from taking advanced mathematics courses in high school, which may limit their future educational attainment (Francis et al. 2019). We move forward with the typical calculation for occupational crowding subject to the caveat of these kinds of educational inequalities.

Finally, frontline industries are defined according to the following six industry groupings: (1) Grocery, Convenience, and Drug Stores, (2) Public Transit, (3) Health care, (4) Trucking, Warehouse, and Postal Service, (5) Building Cleaning Services, and (6) Child Care and Social Services.[3] These industry groupings have been selected in other studies (e.g. Stringer 2020) because they consist of direct-service work that was essential to maintaining communities during the onset of COVID-19 in the United States. Other scholars have used O*NET data to estimate occupational differences in opportunities to socially distance or work remotely (Crowley and Doran 2020; Mongey et al. 2021). Ultimately, workers in the six industries we highlight are considered to have among the lowest opportunities to work remotely or socially distance.

V. OCCUPATIONAL CROWDING RESULTS

We calculate the share of workers employed in a frontline industry in 2020 for each major occupation group, as illustrated in Table 1.

*{Table 1 about here}*

---

[3] These categories include the following industries (and their corresponding 2020+ Census Bureau Industry codes): Grocery and related product merchant wholesalers (4470), Supermarkets and other grocery stores (4971), Convenience stores (4972), Pharmacies and drug stores (5070), and General merchandise stores, including warehouse clubs and supercenters (5391). Rail transportation (6080) and Bus service and urban transit (6180). Truck transportation (6170), Warehousing and storage (6390), and Postal service (6370). Cleaning services to buildings and dwellings (7690). Offices of physicians (7970), Outpatient care centers (8090), Home health care services (8170), Other health care services (8180), General medical and surgical hospitals, and specialty hospitals (8191), Psychiatric and substance abuse hospitals (8192), Nursing care facilities (skilled nursing facilities) (8270), and Residential care facilities, except skilled nursing facilities (8290). Individual and family services (8370), Community food and housing, and emergency services (8380), and Child day care services (8470).

Results from the occupational crowding calculations indicate that immigrant workers are overrepresented in three out of five occupational groups with the most frontline workers. More specifically, in both 2019 and 2020, immigrant workers were crowded in healthcare support, transportation and material moving, and building/grounds cleaning and maintenance occupations. They were underrepresented in community and social service occupations, and proportionally represented in healthcare practitioners and technical occupations. However, within immigrant groups, naturalized citizens were overrepresented among healthcare practitioners and technical occupations while non-citizens were underrepresented. As indicated in Table 1, the common frontline occupation groups in which immigrants were crowded tended to have lower median wages in 2020. Among the five occupation groups with the most frontline workers, healthcare practitioners and technical occupations had the highest median wages at $60,000, and community and social service occupations had the second highest median wages at $41,500.

These results are not especially new. Others have found that immigrant workers were overrepresented in frontline work (Gelatt 2020). However, our contribution is to understand whether and how occupational crowding in frontline work changed at the onset of the pandemic. In Table 1, we show differences in the occupational crowding indices from 2019 to 2020. Using a significance level of 0.10, we conducted one-tailed two-proportion z tests to determine the statistical significance of changes from 2019 to 2020 for both the numerator and the denominator of the index. In other words, statistical significance tests of changes in the indices' numerators were calculated independently of statistical significance tests of changes in the indices' denominators.

Table 1 shows that from 2019 to 2020, immigrants were increasingly crowded in common frontline occupations like healthcare support, community and social service, and transportation and material moving occupations. Results illustrate that both naturalized citizens and non-citizens were increasingly crowded in these occupation groups. As shown in Appendix 2, CPS data also suggest that immigrants were increasingly crowded in healthcare support and community and social services during the onset of the pandemic. The CPS data, however, do not confirm increased occupational crowding in transportation and material moving occupations for immigrants. On the other end of the spectrum, some immigrants were pushed out of farming, fishing, and forestry occupations, as evidenced by both CPS and ACS data, but they still remain heavily crowded in these occupations. Transportation and material moving occupations and farming, fishing, and forestry occupations often have fewer face-to-face interactions than the other three occupation categories, implying that there is less opportunity to contract COVID-19 at work and that immigrants may have been pushed out of these industries for this reason.

Because men and women face different experiences in the labor market, and because many women face burdens of gender discrimination as well as racial and ethnic discrimination, we next examine immigrants' occupational patterns within women's and men's labor markets. We find that among men in the United States, immigrant men are overrepresented in many of the occupation groups in which the largest share of workers are in frontline industries, especially healthcare support and building/grounds cleaning and maintenance. As shown in Table 2, there were some increases in their occupational crowding indices from 2019 to 2020 in these frontline occupations, but the CPS data do not confirm these patterns (as shown in Appendix 2). However, immigrant women were noticeably pushed into frontline work during the pandemic. As shown in Table 3, among all women, immigrant women were already overrepresented in healthcare support and building/grounds cleaning and maintenance (occupation groups where 77% and 46% of workers are in a frontline industry, respectively). However, their crowding index in healthcare support and community and social services occupations increased from 2019 to 2020, a trend which is confirmed by CPS data calculations. These increases in frontline occupational crowding were especially common among women who are not citizens.

*{Table 2 and Table 3 about here}*

Immigrant women were increasingly crowded into frontline work, suggesting they were increasingly exposed to COVID-19 through their work. We posit that this also implies they were used as a mechanism to prop up the health of men and US-born women, who were able to reduce their work-based virus exposure as a result. To support this assertion, we calculate the occupational crowding indices for US-born workers, stratified by race and ethnicity. Specifically, we calculate the occupational crowding indices for major occupation groups for US-born Black workers, US-born White non-Hispanic workers, and US-born Hispanic workers.

*{Table 4 about here}*

Results in Table 4 indicate that in both 2019 and 2020, US-born White non-Hispanic workers were either proportionately represented or underrepresented in occupations that are commonly held by frontline workers. In fact, they were only proportionately represented in the highest-paying occupations common to frontline workers (namely, healthcare practitioners and technical occupations and community and social service occupations). Their occupational crowding indices changed little from 2019 to 2020. Black and Hispanic US-born workers, on the other hand, were generally overrepresented in occupations common to frontline workers (aside from healthcare practitioners and technical occupations, in which they were proportionally represented or underrepresented). Both groups were especially overrepresented in community and social service occupations. From 2019 to 2020, Black US-born workers' occupational crowding indices decreased in many of the common frontline occupations, namely in healthcare occupations and community and social service occupations. This result is in line with earlier research which find that Black workers, and Black women in particular, were among the most likely to have lost their jobs during the pandemic, even in essential work like healthcare and transportation (Holder et al. 2021). Hispanic workers' occupational crowding indices in frontline work decreased less drastically. Overall, trends within gender by race and ethnicity of US-born workers were similar, as illustrated in Table 5 and Table 6. However, among men, Black US-born men were largely crowded out of healthcare occupations.

*{Table 5 and Table 6 about here}*

On the whole, these results do not suggest that immigrants crowded-out racial minorities from dangerous frontline jobs. Instead, we suggest that labor dynamics shifted during the pandemic such that employers had more power to replace the existing labor force with cheaper immigrant workers who had relatively little bargaining power. Ultimately, the crowding of immigrant workers in frontline occupations during COVID-19 can be attributed to a combination of discriminatory and exploitative policies operating at a global scale. Neoliberal policies around the world have created a global circuit of precarious immigrant labor that can be exploited during health or care crises. We have already seen this dynamic play out in the case of global care chains (Fraser, 2017), for example. Now in the case of the pandemic, immigrant workers were in especially precarious positions. Non-citizens were trapped geographically and often had harder time accessing and paying for medical treatment (Page et al 2020). In addition, many non-citizens were ineligible for pandemic relief programs (Suro and Findling 2020). They therefore had little bargaining power and faced high rates of exploitation.

# VI. CONCLUSION

This study suggests that immigrant women were increasingly crowded into frontline occupations during the onset of COVID-19, while native-born White workers were largely able to cluster in safe occupations. Immigrant women faced large increases in crowding in healthcare support occupations, which were among the riskiest occupations during COVID-19.These jobs require high levels of contact and physical proximity at work, and they are less amenable to remote work. These job features thus contribute to increased exposure to COVID-19 and other contagious diseases. This finding is consistent with stratification economists' theories of discrimination which posit that material gains accrue to more privileged economic participants as a result of discrimination against minoritized and marginalized groups. In this case, occupational discrimination against immigrant women allowed native-born workers to shift out of dangerous frontline work. We argue that the occupational crowding hypothesis – that discrimination props up the wages of White/male workers – can be extended to health economics, whereby occupational crowding is used to prop up the health of native-born workers.

Ultimately, these results highlight the exploitation of immigrant workers in the United States, many of whom have consistently supported the country's economic prosperity and faced health risks in doing so, but have received little by way of necessary healthcare support. In the long run, low-pay in essential jobs, in conjunction with health risks and stressful working conditions, will likely contribute to worker burnout, high turnover, and reduced entry into frontline jobs. These effects come into play across the occupational spectrum, including those in care provision, and our results suggest that these adverse effects are disproportionately borne by immigrant workers.

More research is needed on the consequences of greater workplace exposure to contagion among immigrant workers and Black and Hispanic workers, both in the short- and long-term. We also see a need for studies on cultural factors, social norms, and racial/ethnic bias that might affect differences in Black workers' and White workers' vulnerability to infection exposure. Adding an intersectional dimension to this research is critical given the additional biases that Black, Hispanic, or immigrant women may face, and the gender roles which they are expected to perform. Additional research on the experiences of immigrant workers and Black and Hispanic workers in essential jobs will help to inform which workplace supports – including collective bargaining, education and training programs, and stronger care infrastructures – could help to mitigate the health risks they face.

Table 1. Occupational Crowding Index All Immigrants (Non-Citizens and Naturalized Citizens), 2019 and 2020

| Occupation Group | All immigrants | | | Non-Citizens | | | Naturalized Citizens | | | % in frontline industries | Median wage |
|---|---|---|---|---|---|---|---|---|---|---|---|
| | 2019 | 2020 | Difference | 2019 | 2020 | Difference | 2019 | 2020 | Difference | | |
| Healthcare Practitioners and Technical | 0.96 | 0.96 | 0.00 * | 0.54 | 0.54 | 0.01 | 1.19 | 1.18 | -0.01 * | 79% | $ 60,000 |
| Healthcare Support | 1.69 | 1.77 | 0.08 ** | 1.51 | 1.58 | 0.07 ** | 1.77 | 1.84 | 0.07 * | 76% | $ 21,200 |
| Community and Social Service | 0.65 | 0.69 | 0.04 *** | 0.42 | 0.47 | 0.05 * | 0.73 | 0.77 | 0.04 * | 40% | $ 41,500 |
| Transportation and Material Moving | 1.34 | 1.37 | 0.03 *** | 1.89 | 2.01 | 0.12 ** | 1.14 | 1.16 | 0.02 * | 36% | $ 28,000 |
| Building/Grounds Cleaning and Maintenance | 2.33 | 2.28 | -0.05 ** | 4.38 | 4.19 | -0.19 *** | 1.52 | 1.54 | 0.02 * | 33% | $ 15,100 |
| Office and Administrative Support | 0.90 | 0.92 | 0.02 * | 0.65 | 0.69 | 0.04 ** | 1.00 | 1.01 | 0.01 * | 22% | $ 32,700 |
| Personal Care and Service | 1.71 | 1.77 | 0.07 *** | 1.77 | 1.77 | 0.00 | 1.68 | 1.78 | 0.10 *** | 21% | $ 9,000 |
| Sales | 1.07 | 1.08 | 0.01 * | 1.14 | 1.15 | 0.00 | 1.09 | 1.10 | 0.01 * | 19% | $ 30,000 |
| Life, Physical, and Social Science | 1.28 | 1.29 | 0.02 * | 1.59 | 1.43 | -0.15 *** | 1.05 | 1.14 | 0.09 *** | 12% | $ 62,000 |
| Management | 0.93 | 0.93 | 0.00 *** | 0.79 | 0.82 | 0.03 | 0.95 | 0.94 | -0.01 *** | 11% | $ 70,000 |
| Food Preparation and Serving | 1.83 | 1.83 | 0.00 * | 2.29 | 2.33 | 0.04 | 1.26 | 1.28 | 0.01 * | 10% | $ 15,000 |
| Business and Financial Operations | 0.91 | 0.92 | 0.01 * | 0.84 | 0.81 | -0.03 * | 0.99 | 1.00 | 0.01 * | 8% | $ 64,000 |
| Education Instruction and Library | 0.71 | 0.73 | 0.02 * | 0.65 | 0.69 | 0.04 *** | 0.71 | 0.70 | -0.01 * | 7% | $ 41,700 |
| Installation, Maintenance, and Repair | 1.03 | 1.06 | 0.03 ** | 1.12 | 1.01 | -0.11 *** | 1.00 | 1.02 | 0.01 * | 7% | $ 45,000 |
| Computer and Mathematical | 1.69 | 1.68 | -0.01 *** | 2.19 | 2.28 | 0.09 * | 1.44 | 1.43 | -0.01 * | 7% | $ 85,000 |
| Production | 1.55 | 1.54 | 0.00 | 2.46 | 2.49 | 0.03 * | 1.32 | 1.33 | 0.01 * | 5% | $ 34,800 |
| Protective Service | 0.61 | 0.62 | 0.01 * | 0.54 | 0.57 | 0.03 | 0.69 | 0.70 | 0.01 * | 5% | $ 50,000 |
| Arts, Design, Entertainment, Sports, and Media | 0.99 | 0.99 | 0.00 * | 1.14 | 1.19 | 0.05 | 0.89 | 0.86 | -0.03 * | 3% | $ 30,000 |
| Farming, Fishing, and Forestry | 2.90 | 2.61 | -0.29 *** | 6.00 | 5.89 | -0.11 ** | 1.05 | 0.99 | -0.05 * | 2% | $ 20,800 |
| Legal | 0.55 | 0.58 | 0.04 *** | 0.54 | 0.55 | 0.01 * | 0.64 | 0.69 | 0.05 * | 2% | $ 75,000 |
| Architecture and Engineering | 1.28 | 1.31 | 0.03 * | 1.54 | 1.60 | 0.06 *** | 1.30 | 1.31 | 0.01 * | 1% | $ 83,000 |
| Construction and Extraction | 1.73 | 1.71 | -0.02 ** | 2.42 | 2.43 | 0.01 * | 0.95 | 0.94 | -0.01 * | 1% | $ 32,800 |

*=statistically significant change in denominator, **= statistically significant change in numerator, ***= statistically significant change in denominator and numerator

Table 2. Occupational Crowding Index for Non-Citizens and Naturalized Citizens, Among Men, 2019 and 2020

| Occupation Group | All Immigrants | | | Non-Citizens | | | Naturalized Citizens | | | % in frontline industries |
|---|---|---|---|---|---|---|---|---|---|---|
| | 2019 | 2020 | Difference | 2019 | 2020 | Difference | 2019 | 2020 | Difference | |
| Healthcare Practitioners and Technical | 1.28 | 1.08 | -0.20 * | 0.78 | 0.70 | -0.08 * | 1.60 | 1.32 | -0.28 *** | 77% |
| Healthcare Support | 2.04 | 2.07 | 0.03 * | 1.89 | 2.00 | 0.10 | 2.15 | 2.12 | -0.03 * | 76% |
| Transportation and Material Moving | 1.40 | 1.43 | 0.04 * | 1.57 | 1.61 | 0.04 | 1.27 | 1.31 | 0.04 * | 36% |
| Community and Social Service | 0.80 | 0.86 | 0.05 *** | 0.74 | 0.77 | 0.03 * | 0.85 | 0.92 | 0.07 | 25% |
| Building/Grounds Cleaning and Maintenance | 2.00 | 1.98 | -0.02 * | 2.84 | 2.70 | -0.14 ** | 1.34 | 1.39 | 0.06 * | 23% |
| Office and Administrative Support | 1.17 | 1.18 | 0.01 * | 1.01 | 1.03 | 0.02 | 1.28 | 1.29 | 0.01 * | 21% |
| Sales | 1.08 | 1.08 | 0.01 * | 1.07 | 1.10 | 0.04 ** | 1.08 | 1.07 | -0.01 * | 15% |
| Food Preparation and Serving | 2.24 | 2.16 | -0.08 *** | 3.20 | 3.05 | -0.14 ** | 1.55 | 1.50 | -0.05 *** | 8% |
| Life, Physical, and Social Science | 1.22 | 1.25 | 0.03 * | 1.63 | 1.56 | -0.07 | 0.93 | 1.03 | 0.10 *** | 8% |
| Personal Care and Service | 1.69 | 1.80 | 0.11 * | 1.57 | 1.61 | 0.04 | 1.77 | 1.94 | 0.17 * | 7% |
| Management | 0.96 | 0.96 | 0.00 * | 0.95 | 0.96 | 0.01 * | 0.97 | 0.96 | -0.01 | 7% |
| Installation, Maintenance, and Repair | 1.11 | 1.15 | 0.04 * | 1.09 | 1.15 | 0.07 ** | 1.13 | 1.14 | 0.01 * | 7% |
| Computer and Mathematical | 1.64 | 1.61 | -0.04 *** | 1.99 | 1.96 | -0.03 *** | 1.41 | 1.37 | -0.04 ** | 6% |
| Business and Financial Operations | 0.85 | 0.89 | 0.04 *** | 0.74 | 0.79 | 0.05 *** | 0.93 | 0.95 | 0.02 | 5% |
| Production | 1.38 | 1.38 | 0.01 * | 1.58 | 1.54 | -0.03 | 1.22 | 1.25 | 0.04 * | 4% |
| Protective Service | 0.62 | 0.63 | 0.01 * | 0.46 | 0.47 | 0.01 | 0.71 | 0.72 | 0.01 * | 4% |
| Education Instruction and Library | 0.79 | 0.83 | 0.04 *** | 0.96 | 1.06 | 0.10 ** | 0.67 | 0.67 | 0.00 * | 2% |
| Arts, Design, Entertainment, Sports, and Media | 0.95 | 0.96 | 0.01 * | 1.07 | 1.12 | 0.05 * | 0.87 | 0.85 | -0.02 | 2% |
| Legal | 0.42 | 0.49 | 0.07 *** | 0.38 | 0.38 | 0.00 | 0.44 | 0.54 | 0.10 * | 1% |
| Architecture and Engineering | 1.21 | 1.22 | 0.01 * | 1.10 | 1.17 | 0.07 *** | 1.28 | 1.25 | -0.03 | 1% |
| Construction and Extraction | 1.87 | 1.86 | -0.01 * | 2.91 | 2.85 | -0.06 | 1.06 | 1.07 | 0.01 * | 1% |
| Farming, Fishing, and Forestry | 2.82 | 2.42 | -0.40 *** | 4.85 | 4.08 | -0.76 ** | 0.99 | 0.88 | -0.10 *** | 1% |

*=statistically significant change in denominator, **= statistically significant change in numerator, ***= statistically significant change in denominator and numerator

Table 3. Occupational Crowding Index for Non-Citizens and Naturalized Citizens, Among Women, 2019 and 2020

| Occupation Group | All Immigrants | | | Non-Citizens | | | Naturalized Citizens | | | % in frontline industries |
|---|---|---|---|---|---|---|---|---|---|---|
| | 2019 | 2020 | Difference | 2019 | 2020 | Difference | 2019 | 2020 | Difference | |
| Healthcare Practitioners and Technical | 0.87 | 0.87 | 0.01 | 0.54 | 0.54 | 0.01 | 1.05 | 1.06 | 0.01 | 79% |
| Healthcare Support | 1.56 | 1.63 | 0.07 *** | 1.51 | 1.58 | 0.07 ** | 1.59 | 1.66 | 0.07 * | 77% |
| Community and Social Service | 0.57 | 0.61 | 0.04 ** | 0.42 | 0.47 | 0.05 ** | 0.67 | 0.70 | 0.03 | 47% |
| Building/Grounds Cleaning and Maintenance | 2.87 | 2.78 | -0.08 *** | 4.38 | 4.19 | -0.19 | 1.82 | 1.79 | -0.03 * | 46% |
| Transportation and Material Moving | 1.38 | 1.41 | 0.04 * | 1.89 | 2.01 | 0.12 *** | 1.06 | 1.03 | -0.02 * | 35% |
| Personal Care and Service | 1.63 | 1.68 | 0.04 | 1.77 | 1.77 | 0.00 | 1.55 | 1.62 | 0.07 | 25% |
| Sales | 1.08 | 1.09 | 0.02 | 1.14 | 1.15 | 0.00 | 1.03 | 1.06 | 0.03 * | 24% |
| Office and Administrative Support | 0.79 | 0.80 | 0.02 *** | 0.65 | 0.69 | 0.04 ** | 0.87 | 0.88 | 0.00 * | 22% |
| Management | 0.89 | 0.89 | 0.00 * | 0.79 | 0.82 | 0.03 | 0.95 | 0.94 | -0.01 * | 17% |
| Life, Physical, and Social Science | 1.33 | 1.33 | 0.00 * | 1.59 | 1.43 | -0.15 ** | 1.17 | 1.26 | 0.09 * | 16% |
| Food Preparation and Serving | 1.54 | 1.59 | 0.05 ** | 2.29 | 2.33 | 0.04 | 1.07 | 1.11 | 0.05 | 12% |
| Installation, Maintenance, and Repair | 0.91 | 1.02 | 0.12 | 1.12 | 1.01 | -0.11 | 0.77 | 1.03 | 0.26 | 10% |
| Education Instruction and Library | 0.68 | 0.69 | 0.01 | 0.65 | 0.69 | 0.04 ** | 0.70 | 0.69 | -0.01 | 9% |
| Business and Financial Operations | 0.96 | 0.96 | -0.01 *** | 0.84 | 0.81 | -0.03 | 1.04 | 1.04 | 0.00 * | 9% |
| Computer and Mathematical | 1.79 | 1.85 | 0.07 * | 2.19 | 2.28 | 0.09 * | 1.56 | 1.60 | 0.05 | 9% |
| Production | 1.90 | 1.90 | 0.01 | 2.46 | 2.49 | 0.03 | 1.54 | 1.53 | -0.01 | 8% |
| Protective Service | 0.66 | 0.68 | 0.01 | 0.54 | 0.57 | 0.03 | 0.73 | 0.73 | 0.00 | 6% |
| Farming, Fishing, and Forestry | 3.22 | 3.22 | 0.00 * | 6.00 | 5.89 | -0.11 * | 1.23 | 1.30 | 0.07 * | 5% |
| Arts, Design, Entertainment, Sports, and Media | 0.96 | 0.96 | 0.00 * | 1.14 | 1.19 | 0.05 * | 0.85 | 0.83 | -0.02 * | 4% |
| Construction and Extraction | 1.47 | 1.41 | -0.07 * | 2.42 | 2.43 | 0.01 | 0.87 | 0.75 | -0.12 * | 2% |
| Legal | 0.67 | 0.69 | 0.02 | 0.54 | 0.55 | 0.01 | 0.74 | 0.78 | 0.04 | 2% |
| Architecture and Engineering | 1.49 | 1.60 | 0.10 ** | 1.54 | 1.60 | 0.06 | 1.47 | 1.59 | 0.13 | 2% |

*=statistically significant change in denominator, ** = statistically significant change in numerator, ***= statistically significant change in denominator and numerator

Table 4. Occupational Crowding Index U.S. Born Workers, by Race and Ethnicity, 2019 and 2020

| Occupation Group | Black | | | White, non-Hispanic | | | Hispanic | | | % in frontline industries |
|---|---|---|---|---|---|---|---|---|---|---|
| | 2019 | 2020 | Difference | 2019 | 2020 | Difference | 2019 | 2020 | Difference | |
| Healthcare Practitioners and Technical | 1.00 | 0.92 | -0.07 *** | 1.00 | 1.01 | 0.01 *** | 0.95 | 0.96 | 0.01 *** | 79% |
| Healthcare Support | 1.64 | 1.57 | -0.07 *** | 0.76 | 0.75 | 0.00 *** | 1.27 | 1.20 | -0.07 * | 76% |
| Community and Social Service | 2.20 | 2.14 | -0.06 *** | 0.95 | 0.94 | -0.01 *** | 1.61 | 1.58 | -0.03 *** | 40% |
| Transportation and Material Moving | 1.50 | 1.56 | 0.06 *** | 0.86 | 0.84 | -0.02 *** | 1.17 | 1.18 | 0.01 *** | 36% |
| Building/Grounds Cleaning and Maintenance | 1.10 | 1.10 | 0.00 * | 0.75 | 0.75 | 0.00 *** | 1.04 | 0.98 | -0.06 *** | 33% |
| Office and Administrative Support | 1.07 | 1.02 | -0.05 *** | 0.99 | 0.99 | 0.00 *** | 1.23 | 1.22 | -0.01 *** | 22% |
| Personal Care and Service | 0.93 | 0.95 | 0.02 * | 0.88 | 0.86 | -0.02 *** | 1.00 | 1.05 | 0.04 *** | 21% |
| Sales | 0.72 | 0.71 | -0.01 * | 1.03 | 1.03 | 0.00 *** | 1.02 | 1.02 | 0.00 *** | 19% |
| Life, Physical, and Social Science | 0.68 | 0.67 | -0.01 * | 0.94 | 0.94 | 0.00 *** | 1.01 | 0.99 | -0.02 * | 12% |
| Management | 0.66 | 0.72 | 0.05 *** | 1.07 | 1.07 | 0.00 *** | 0.87 | 0.87 | 0.00 *** | 11% |
| Food Preparation and Serving | 1.22 | 1.24 | 0.02 * | 0.79 | 0.77 | -0.02 *** | 1.17 | 1.23 | 0.05 *** | 10% |
| Business and Financial Operations | 1.09 | 1.10 | 0.01 *** | 1.00 | 1.00 | 0.00 *** | 1.10 | 1.09 | -0.01 *** | 8% |
| Education Instruction and Library | 1.19 | 1.18 | -0.02 *** | 1.04 | 1.04 | 0.00 *** | 1.27 | 1.25 | -0.02 *** | 7% |
| Installation, Maintenance, and Repair | 0.52 | 0.53 | 0.01 * | 1.09 | 1.09 | 0.00 *** | 0.91 | 0.92 | 0.01 *** | 7% |
| Computer and Mathematical | 0.83 | 0.83 | 0.00 * | 0.86 | 0.86 | -0.01 *** | 0.85 | 0.89 | 0.04 *** | 7% |
| Production | 0.94 | 0.91 | -0.03 * | 0.95 | 0.96 | 0.01 *** | 0.81 | 0.82 | 0.01 *** | 5% |
| Protective Service | 1.76 | 1.76 | 0.00 * | 0.95 | 0.94 | -0.01 *** | 1.39 | 1.39 | 0.00 * | 5% |
| Arts, Design, Entertainment, Sports, and Media | 0.62 | 0.67 | 0.05 * | 1.04 | 1.03 | -0.01 *** | 0.95 | 1.00 | 0.06 *** | 3% |
| Farming, Fishing, and Forestry | 0.31 | 0.32 | 0.01 * | 0.76 | 0.82 | 0.06 *** | 0.93 | 0.88 | -0.05 * | 2% |
| Legal | 0.90 | 0.90 | 0.00 * | 1.09 | 1.08 | -0.01 *** | 1.25 | 1.26 | 0.01 *** | 2% |
| Architecture and Engineering | 0.54 | 0.51 | -0.03 *** | 0.99 | 0.98 | -0.01 *** | 0.83 | 0.91 | 0.07 *** | 1% |
| Construction and Extraction | 0.49 | 0.56 | 0.07 *** | 0.96 | 0.96 | 0.00 *** | 1.01 | 0.98 | -0.04 * | 1% |

*=statistically significant change in denominator, **= statistically significant change in numerator, ***= statistically significant change in denominator and numerator

Table 5. Occupational Crowding Index U.S. Born Men, by Race and Ethnicity, 2019 and 2020

| Occupation Group | Black | | | White, non-Hispanic | | | Hispanic | | | % in frontline industries |
|---|---|---|---|---|---|---|---|---|---|---|
| | 2019 | 2020 | Difference | 2019 | 2020 | Difference | 2019 | 2020 | Difference | |
| Healthcare Practitioners and Technical | 0.89 | 0.66 | -0.23 *** | 0.91 | 0.78 | -0.12 *** | 1.10 | 0.93 | -0.17 * | 77% |
| Healthcare Support | 1.49 | 1.48 | -0.01 * | 0.68 | 0.69 | 0.01 * | 1.43 | 1.27 | -0.15 *** | 76% |
| Transportation and Material Moving | 1.51 | 1.53 | 0.02 *** | 0.86 | 0.84 | -0.02 *** | 1.18 | 1.20 | 0.02 *** | 36% |
| Community and Social Service | 2.30 | 2.21 | -0.09 | 0.94 | 0.94 | 0.00 *** | 0.62 | 0.60 | -0.02 * | 25% |
| Building/Grounds Cleaning and Maintenance | 1.20 | 1.19 | -0.01 *** | 0.81 | 0.80 | -0.01 *** | 1.09 | 1.04 | -0.04 * | 23% |
| Office and Administrative Support | 1.21 | 1.08 | -0.13 *** | 0.90 | 0.90 | 0.00 *** | 1.33 | 1.36 | 0.03 *** | 21% |
| Sales | 0.64 | 0.64 | 0.00 *** | 1.04 | 1.05 | 0.01 *** | 0.97 | 0.98 | 0.01 *** | 15% |
| Food Preparation and Serving | 1.51 | 1.51 | 0.00 *** | 0.67 | 0.67 | -0.01 *** | 1.37 | 1.41 | 0.05 *** | 8% |
| Life, Physical, and Social Science | 0.71 | 0.70 | -0.02 * | 0.96 | 0.95 | -0.01 *** | 0.99 | 0.90 | -0.09 * | 8% |
| Personal Care and Service | 1.34 | 1.36 | 0.02 *** | 0.80 | 0.77 | -0.03 *** | 1.20 | 1.27 | 0.07 *** | 7% |
| Management | 0.58 | 0.59 | 0.00 | 1.08 | 1.08 | 0.01 *** | 0.80 | 0.80 | 0.00 *** | 7% |
| Installation, Maintenance, and Repair | 0.50 | 0.50 | 0.00 *** | 1.08 | 1.08 | 0.00 *** | 0.90 | 0.92 | 0.02 *** | 7% |
| Computer and Mathematical | 0.79 | 0.81 | 0.02 | 0.87 | 0.87 | 0.00 *** | 0.88 | 0.93 | 0.05 *** | 6% |
| Business and Financial Operations | 0.95 | 0.97 | 0.01 | 1.03 | 1.02 | 0.00 *** | 1.09 | 1.04 | -0.05 * | 5% |
| Production | 0.85 | 0.83 | -0.02 *** | 1.00 | 1.01 | 0.01 *** | 0.81 | 0.81 | -0.01 * | 4% |
| Protective Service | 1.56 | 1.51 | -0.05 * | 0.97 | 0.97 | -0.01 *** | 1.40 | 1.39 | -0.01 * | 4% |
| Education Instruction and Library | 1.39 | 1.35 | -0.05 * | 1.01 | 1.00 | -0.01 *** | 1.41 | 1.35 | -0.07 * | 2% |
| Arts, Design, Entertainment, Sports, and Media | 0.83 | 0.86 | 0.03 | 1.01 | 1.00 | -0.01 *** | 1.03 | 1.08 | 0.06 *** | 2% |
| Legal | 1.27 | 1.22 | -0.05 * | 1.14 | 1.13 | -0.01 *** | 1.30 | 1.20 | -0.10 | 1% |
| Architecture and Engineering | 0.57 | 0.56 | -0.01 | 1.00 | 0.99 | -0.02 *** | 0.83 | 0.96 | 0.12 *** | 1% |
| Construction and Extraction | 0.49 | 0.53 | 0.04 *** | 0.95 | 0.95 | 0.01 *** | 1.01 | 0.97 | -0.04 * | 1% |
| Farming, Fishing, and Forestry | 0.31 | 0.34 | 0.04 *** | 0.80 | 0.88 | 0.08 *** | 0.93 | 0.82 | -0.10 *** | 1% |

*=statistically significant change in denominator, **= statistically significant change in numerator, ***= statistically significant change in denominator and numerator

| Occupation Group | Black | | | White, non-Hispanic | | | Hispanic | | | % in frontline industries |
|---|---|---|---|---|---|---|---|---|---|---|
| | 2019 | 2020 | Difference | 2019 | 2020 | Difference | 2019 | 2020 | Difference | |
| Healthcare Practitioners and Technical | 0.95 | 0.89 | -0.05 ** | 1.04 | 1.04 | 0.00 *** | 0.89 | 0.90 | 0.02 *** | 79% |
| Healthcare Support | 1.63 | 1.62 | -0.01 *** | 0.78 | 0.77 | -0.01 *** | 1.25 | 1.19 | -0.06 * | 77% |
| Community and Social Service | 2.05 | 2.00 | -0.05 ** | 0.96 | 0.95 | -0.01 *** | 1.63 | 1.62 | -0.01 *** | 47% |
| Building/Grounds Cleaning and Maintenance | 0.98 | 0.98 | 0.00 * | 0.64 | 0.66 | 0.02 * | 1.00 | 0.91 | -0.09 *** | 46% |
| Transportation and Material Moving | 1.56 | 1.56 | 0.01 *** | 0.84 | 0.83 | -0.01 *** | 1.13 | 1.12 | -0.01 * | 35% |
| Personal Care and Service | 0.81 | 0.86 | 0.05 | 0.91 | 0.89 | -0.02 *** | 0.95 | 0.99 | 0.04 *** | 25% |
| Sales | 0.85 | 0.84 | -0.01 *** | 1.00 | 0.99 | 0.00 *** | 1.09 | 1.08 | 0.00 *** | 24% |
| Office and Administrative Support | 1.01 | 1.02 | 0.01 *** | 1.02 | 1.02 | 0.00 *** | 1.20 | 1.17 | -0.03 * | 22% |
| Management | 0.79 | 0.89 | 0.10 *** | 1.05 | 1.04 | -0.01 *** | 0.97 | 0.96 | -0.01 * | 17% |
| Life, Physical, and Social Science | 0.69 | 0.68 | -0.01 * | 0.93 | 0.93 | 0.00 *** | 1.04 | 1.08 | 0.05 *** | 16% |
| Food Preparation and Serving | 1.01 | 1.04 | 0.02 | 0.88 | 0.86 | -0.02 *** | 1.03 | 1.08 | 0.05 *** | 12% |
| Installation, Maintenance, and Repair | 1.21 | 1.04 | -0.17 ** | 0.96 | 0.99 | 0.02 * | 1.18 | 0.89 | -0.29 *** | 10% |
| Education Instruction and Library | 1.07 | 1.06 | -0.01 *** | 1.07 | 1.06 | 0.00 *** | 1.20 | 1.20 | 0.00 *** | 9% |
| Business and Financial Operations | 1.17 | 1.17 | 0.00 *** | 0.98 | 0.98 | 0.00 *** | 1.11 | 1.12 | 0.02 *** | 9% |
| Computer and Mathematical | 1.11 | 1.06 | -0.06 ** | 0.82 | 0.80 | -0.02 *** | 0.84 | 0.87 | 0.04 *** | 9% |
| Production | 1.35 | 1.24 | -0.10 ** | 0.80 | 0.81 | 0.01 *** | 0.88 | 0.93 | 0.05 *** | 8% |
| Protective Service | 2.71 | 2.72 | 0.00 | 0.80 | 0.80 | 0.00 *** | 1.41 | 1.42 | 0.01 * | 6% |
| Farming, Fishing, and Forestry | 0.35 | 0.24 | -0.12 *** | 0.62 | 0.62 | 0.00 *** | 1.02 | 1.15 | 0.13 *** | 5% |
| Arts, Design, Entertainment, Sports, and Media | 0.55 | 0.61 | 0.06 * | 1.04 | 1.03 | -0.01 *** | 1.00 | 1.05 | 0.05 *** | 4% |
| Construction and Extraction | 0.62 | 0.88 | 0.26 *** | 0.94 | 0.92 | -0.02 *** | 1.01 | 1.14 | 0.14 *** | 2% |
| Legal | 0.91 | 0.91 | 0.00 | 1.05 | 1.03 | -0.02 *** | 1.25 | 1.34 | 0.09 *** | 2% |
| Architecture and Engineering | 0.79 | 0.60 | -0.19 ** | 0.88 | 0.88 | 0.00 ** | 1.04 | 0.86 | -0.18 ** | 2% |

*=statistically significant change in denominator, **= statistically significant change in numerator, ***= statistically significant change in denominator and numerator

**APPENDIX 1. Country/Continent of Birth for Immigrants in the Labor Force**

Here we include descriptive statistics on the birthplaces of immigrant workers in our sample. More specifically, Appendix 1 Table 1 illustrates the indicated place of birth of immigrants in the labor force who were sampled during the 2019 or 2020 American Community Surveys. Note that, some respondents' country of birth is not available, and their general continent or region is provided instead, which makes disaggregating data by country of birth imprecise. Still, Appendix 1 Table 1 offers a more detailed picture of the immigrant workers we discuss in this paper.

Many of the immigrants in our sample (22.50%) indicated they were born in Mexico, which was even more common when honing the sample to non-citizens (31.88%). Other immigrants commonly indicated they were born in India (9.29%) or China (7.45%). A decent share were born is South America (7.68%) or Central America (7.13%), but their specific country of birth was not observable in the public-use data.

Appendix 1. Table 1. Birthplaces of Immigrants in the Laborforce, from 2019 and 2020 American Community Surveys

| Reported Region | All Immigrants | Non-Citizens | Naturalized Citizens |
|---|---|---|---|
| Mexico | 22.50% | 31.88% | 15.22% |
| India | 9.29% | 9.41% | 9.20% |
| South America (country not specified) | 7.68% | 7.12% | 8.11% |
| China | 7.45% | 6.81% | 7.95% |
| Central America (country not specified) | 7.13% | 9.57% | 5.24% |
| West Indies (country not specified) | 5.92% | 4.33% | 7.15% |
| Philippines | 5.41% | 3.24% | 7.09% |
| Africa (country not specified) | 4.89% | 3.95% | 5.62% |
| Vietnam | 3.75% | 1.57% | 5.44% |
| Other USSR/Russia (country not specified) | 2.89% | 1.69% | 3.83% |
| Korea | 2.56% | 1.89% | 3.07% |
| Cuba | 2.38% | 2.02% | 2.66% |
| Canada | 1.96% | 2.24% | 1.74% |
| Poland | 1.10% | 0.64% | 1.46% |
| Germany | 1.03% | 1.16% | 0.93% |
| Iran | 0.99% | 0.56% | 1.32% |
| United Kingdom (country not specified) | 0.93% | 1.12% | 0.79% |
| Japan | 0.74% | 1.22% | 0.36% |
| England | 0.71% | 0.71% | 0.71% |
| Yugoslavia | 0.61% | 0.31% | 0.85% |
| Thailand | 0.60% | 0.44% | 0.72% |
| Italy | 0.54% | 0.50% | 0.57% |
| France | 0.48% | 0.58% | 0.39% |
| Romania | 0.47% | 0.22% | 0.66% |
| Laos | 0.41% | 0.15% | 0.61% |
| Cambodia (Kampuchea) | 0.40% | 0.13% | 0.61% |
| Australia and New Zealand | 0.40% | 0.58% | 0.27% |
| Iraq | 0.38% | 0.22% | 0.51% |
| Portugal | 0.37% | 0.24% | 0.47% |
| Nepal | 0.36% | 0.43% | 0.31% |
| Turkey | 0.35% | 0.34% | 0.35% |
| Israel/Palestine | 0.32% | 0.21% | 0.40% |
| Ireland | 0.30% | 0.27% | 0.33% |
| Lebanon | 0.30% | 0.13% | 0.44% |
| Indonesia | 0.27% | 0.26% | 0.29% |
| Greece | 0.26% | 0.18% | 0.32% |
| Spain | 0.26% | 0.32% | 0.21% |
| Albania | 0.25% | 0.16% | 0.32% |
| Pacific Islands (country not specified) | 0.25% | 0.29% | 0.22% |
| Bulgaria | 0.24% | 0.13% | 0.32% |
| Afghanistan | 0.21% | 0.21% | 0.21% |
| Malaysia | 0.19% | 0.21% | 0.18% |
| Syria | 0.19% | 0.13% | 0.23% |
| Netherlands | 0.18% | 0.22% | 0.15% |
| Jordan | 0.17% | 0.11% | 0.22% |
| Asia (country not specified) | 0.17% | 0.14% | 0.20% |
| Czechoslovakia | 0.16% | 0.13% | 0.18% |
| Sweden | 0.12% | 0.14% | 0.10% |
| Hungary | 0.12% | 0.11% | 0.13% |
| Switzerland | 0.11% | 0.12% | 0.09% |
| Atlantic Islands (country not specified) | 0.10% | 0.07% | 0.13% |
| Scotland | 0.10% | 0.10% | 0.09% |
| Singapore | 0.09% | 0.13% | 0.05% |
| Kuwait | 0.09% | 0.05% | 0.11% |
| Saudi Arabia | 0.09% | 0.09% | 0.09% |
| Belgium | 0.08% | 0.10% | 0.07% |
| Lithuania | 0.08% | 0.08% | 0.09% |
| Europe (country not specified) | 0.08% | 0.06% | 0.08% |
| Yemen Arab Republic (North) | 0.08% | 0.05% | 0.11% |
| American Samoa | 0.07% | 0.06% | 0.08% |
| Denmark | 0.07% | 0.09% | 0.05% |
| Austria | 0.06% | 0.09% | 0.04% |
| Finland | 0.05% | 0.06% | 0.04% |
| Latvia | 0.05% | 0.03% | 0.06% |
| United Arab Emirates | 0.05% | 0.04% | 0.05% |
| Americas (country not specified) | 0.04% | 0.04% | 0.04% |
| Norway | 0.04% | 0.07% | 0.02% |
| Other | 0.03% | 0.04% | 0.02% |
| Iceland | 0.01% | 0.01% | 0.01% |
| n | 363,841 | 159,136 | 204,705 |

**APPENDIX 2. Occupational Crowding Indices using Current Populations Survey Data**

In this appendix, we offer a robustness check of our occupational crowding index calculations using CPS data rather than ACS data. We use the same methods of calculation as described in the body of the paper and incorporate the same sample selection criteria (e.g. age). There are, however, two main differences. The first is in our measure of occupation. When using ACS data, our major occupation groups at the four-digit level of detail are consistent with the Standard Occupational Classification system. Because this measure is not available in the CPS data, we use a similar harmonized occupation coding scheme based on the Census Bureau's 2010 occupation classification scheme. We classify major occupations groups according to the broader occupation groups schema suggested by the Census.

The second difference in our CPS estimates has to do with years of coverage. Because annual CPS samples are smaller than that of the ACS, we use several more years of CPS data to calculate the crowding indices before and after the onset of COVID-19. We use observations from 2018, 2019, and January and February of 2020 as the "pre-pandemic" dataset and observations from March 2020 to March 2021 as the "during pandemic" dataset. Note that this means we may also be capturing different occupational crowding dynamics, as occupation-based exposure varied from the first-wave of COVID-19 to the second-wave (Magnusson et al. 2021).

Aside from these two differences, the occupational crowding indices are calculated identically to those in the paper. Results from these calculations are listed in Appendix 2 Table 1. Results among men are in Appendix 2 Table 2 and results among women are in Appendix 2 Table 3.

Appendix 2. Table 1. Occupational Crowding Index using CPS Data, Before and During COVID

| Occupation Group | Immigrants | | | U.S. Born Workers | | | | | | | | | % in frontline industries |
|---|---|---|---|---|---|---|---|---|---|---|---|---|---|
| | | | | Black | | | White Non-Hispanic | | | Hispanic | | | |
| | Pre-COVID | During COVID | Difference | Pre-COVID | During COVID | Difference | Pre-COVID | During COVID | Difference | Pre-COVID | During COVID | Difference | |
| Healthcare Support | 1.48 | 1.51 | 0.03 | 1.77 | 1.75 | -0.01 * | 0.78 | 0.76 | -0.02 *** | 1.27 | 1.33 | 0.05 *** | 67.3% |
| Healthcare Practitioners and Technicians | 0.87 | 0.85 | -0.02 *** | 1.11 | 1.05 | -0.07 *** | 1.01 | 1.02 | 0.01 * | 1.06 | 1.03 | -0.04 *** | 57.6% |
| Community and Social Services | 0.56 | 0.59 | 0.03 * | 2.24 | 2.01 | -0.24 *** | 0.94 | 0.96 | 0.02 * | 1.61 | 1.59 | -0.02 *** | 39.8% |
| Personal Care and Service | 1.71 | 1.71 | 0.00 | 1.24 | 1.15 | -0.09 *** | 0.82 | 0.82 | 0.00 *** | 1.10 | 1.16 | 0.06 *** | 35.8% |
| Transportation and Material Moving | 1.52 | 1.48 | -0.03 ** | 1.48 | 1.53 | 0.05 *** | 0.83 | 0.82 | -0.01 *** | 1.12 | 1.13 | 0.02 *** | 33.5% |
| Building and Grounds Cleaning and Maintenance | 2.87 | 2.86 | -0.02 * | 0.95 | 0.91 | -0.04 * | 0.69 | 0.69 | 0.01 * | 0.91 | 0.87 | -0.04 * | 32.4% |
| Office and Administrative Support | 0.81 | 0.83 | 0.03 ** | 1.11 | 1.13 | 0.01 *** | 0.99 | 0.98 | -0.01 *** | 1.24 | 1.24 | 0.00 *** | 20.0% |
| Sales and Related | 0.96 | 0.97 | 0.01 | 0.76 | 0.75 | -0.01 * | 1.04 | 1.04 | 0.00 *** | 1.06 | 1.02 | -0.04 * | 10.9% |
| Management in Business, Science, and Arts | 0.89 | 0.88 | -0.02 ** | 0.65 | 0.71 | 0.06 ** | 1.08 | 1.08 | 0.00 ** | 0.84 | 0.84 | 0.00 ** | 9.4% |
| Life, Physical, and Social Science | 1.29 | 1.27 | -0.02 * | 0.67 | 0.57 | -0.10 *** | 0.96 | 0.96 | 0.00 * | 1.07 | 1.07 | 0.00 *** | 8.0% |
| Business Operations Specialists | 0.80 | 0.87 | 0.07 ** | 0.99 | 0.98 | -0.01 | 1.03 | 1.02 | -0.01 *** | 0.98 | 1.05 | 0.06 *** | 7.8% |
| Food Preparation and Serving | 2.11 | 2.11 | 0.00 | 1.18 | 1.16 | -0.02 * | 0.74 | 0.74 | 0.00 * | 1.16 | 1.16 | 0.00 *** | 7.2% |
| Installation, Maintenance, and Repair | 0.96 | 1.02 | 0.06 ** | 0.68 | 0.59 | -0.09 *** | 1.06 | 1.05 | -0.01 *** | 1.05 | 1.06 | 0.01 *** | 6.7% |
| Education, Training, and Library | 0.64 | 0.65 | 0.01 * | 1.14 | 1.12 | -0.03 * | 1.06 | 1.06 | 0.01 *** | 1.27 | 1.19 | -0.08 * | 6.3% |
| Computer and Mathematical | 1.68 | 1.72 | 0.04 * | 0.85 | 0.85 | -0.01 * | 0.87 | 0.85 | -0.02 *** | 0.82 | 0.86 | 0.05 *** | 5.1% |
| Financial Specialists | 0.88 | 0.87 | -0.02 *** | 1.00 | 1.03 | 0.03 * | 1.01 | 1.01 | -0.01 *** | 1.02 | 1.15 | 0.13 *** | 4.6% |
| Production | 1.66 | 1.64 | -0.02 | 1.05 | 1.07 | 0.02 *** | 0.89 | 0.89 | 0.00 *** | 0.95 | 0.95 | 0.00 *** | 3.6% |
| Protective Service | 0.51 | 0.45 | -0.06 ** | 1.55 | 1.59 | 0.04 *** | 0.98 | 0.97 | -0.01 *** | 1.28 | 1.35 | 0.07 *** | 2.5% |
| Technicians | 0.77 | 0.83 | 0.06 | 0.79 | 0.66 | -0.13 *** | 1.08 | 1.07 | -0.01 *** | 0.90 | 0.95 | 0.05 *** | 2.5% |
| Farming, Fisheries, and Forestry | 2.89 | 2.60 | -0.29 *** | 0.29 | 0.31 | 0.01 * | 0.73 | 0.80 | 0.06 *** | 0.95 | 0.91 | -0.04 * | 2.4% |
| Arts, Design, Entertainment, Sports, and Media | 0.84 | 0.83 | -0.01 * | 0.76 | 0.84 | 0.08 *** | 1.03 | 1.04 | 0.01 * | 1.19 | 1.09 | -0.11 * | 2.3% |
| Legal | 0.49 | 0.48 | -0.01 * | 0.94 | 0.97 | 0.04 * | 1.10 | 1.09 | 0.00 *** | 1.28 | 1.27 | -0.01 *** | 1.3% |
| Architecture and Engineering | 1.17 | 1.25 | 0.08 *** | 0.53 | 0.46 | -0.06 *** | 1.02 | 1.00 | -0.02 *** | 0.85 | 0.93 | 0.08 *** | 1.2% |
| Construction | 2.14 | 2.20 | 0.05 *** | 0.45 | 0.43 | -0.02 * | 0.89 | 0.90 | 0.00 *** | 0.99 | 0.97 | -0.02 *** | 0.8% |
| Extraction | 0.64 | 0.65 | 0.01 * | 0.28 | 0.40 | 0.11 *** | 1.16 | 1.21 | 0.05 *** | 1.24 | 0.79 | -0.45 *** | 0.1% |

*=statistically significant change in denominator, **= statistically significant change in numerator, ***= statistically significant change in denominator and numerator

Appendix 2. Table 2. Occupational Crowding Index for Men using CPS Data, Before and During COVID

| Occupation Group | Immigrants | | | U.S. Born Workers | | | | | | | | | % in frontline industries |
| --- | --- | --- | --- | --- | --- | --- | --- | --- | --- | --- | --- | --- | --- |
| | | | | Black | | | White Non-Hispanic | | | Hispanic | | | |
| | Pre-COVID | During COVID | Difference | Pre-COVID | During COVID | Difference | Pre-COVID | During COVID | Difference | Pre-COVID | During COVID | Difference | |
| Healthcare Support | 1.78 | 1.81 | 0.03 * | 1.64 | 1.46 | -0.18 *** | 0.75 | 0.73 | -0.01 * | 1.40 | 1.17 | -0.24 *** | 63.5% |
| Healthcare Practitioners and Technicians | 1.01 | 1.05 | 0.04 * | 1.00 | 0.95 | -0.05 | 0.96 | 0.95 | -0.01 *** | 1.38 | 1.20 | -0.18 * | 57.4% |
| Transportation and Material Moving | 1.56 | 1.51 | -0.05 *** | 1.54 | 1.63 | 0.09 *** | 0.83 | 0.82 | -0.01 *** | 1.23 | 1.09 | -0.14 *** | 34.9% |
| Community and Social Services | 0.72 | 0.73 | 0.01 * | 2.25 | 2.02 | -0.23 ** | 0.94 | 0.96 | 0.02 * | 1.59 | 1.23 | -0.36 *** | 26.5% |
| Building and Grounds Cleaning and Maintenance | 2.40 | 2.40 | 0.00 | 1.05 | 1.03 | -0.02 * | 0.77 | 0.77 | 0.00 * | 0.98 | 0.91 | -0.07 * | 23.1% |
| Office and Administrative Support | 1.08 | 1.10 | 0.03 * | 1.29 | 1.27 | -0.03 * | 0.91 | 0.90 | -0.01 *** | 1.41 | 1.21 | -0.19 *** | 20.6% |
| Personal Care and Service | 1.76 | 1.83 | 0.06 * | 1.36 | 1.36 | 0.00 * | 0.77 | 0.74 | -0.03 ** | 1.51 | 1.26 | -0.25 *** | 19.5% |
| Sales and Related | 0.98 | 1.00 | 0.02 * | 0.65 | 0.64 | -0.01 * | 1.06 | 1.06 | 0.00 *** | 1.03 | 0.93 | -0.10 *** | 9.4% |
| Installation, Maintenance, and Repair | 1.00 | 1.08 | 0.08 *** | 0.71 | 0.61 | -0.10 *** | 1.04 | 1.03 | -0.01 *** | 1.15 | 1.01 | -0.14 *** | 6.6% |
| Management in Business, Science, and Arts | 0.94 | 0.94 | 0.00 * | 0.57 | 0.62 | 0.05 ** | 1.08 | 1.08 | 0.00 *** | 0.84 | 0.75 | -0.10 *** | 6.1% |
| Business Operations Specialists | 0.75 | 0.85 | 0.10 *** | 1.10 | 1.03 | -0.07 ** | 1.03 | 1.01 | -0.02 *** | 1.31 | 0.94 | -0.38 *** | 5.4% |
| Food Preparation and Serving | 2.53 | 2.49 | -0.04 *** | 1.26 | 1.27 | 0.01 *** | 0.65 | 0.65 | 0.00 * | 1.44 | 1.21 | -0.23 *** | 5.1% |
| Life, Physical, and Social Science | 1.28 | 1.32 | 0.05 * | 0.66 | 0.46 | -0.20 ** | 0.96 | 0.96 | 0.00 * | 1.22 | 0.81 | -0.41 *** | 4.9% |
| Computer and Mathematical | 1.63 | 1.64 | 0.00 *** | 0.80 | 0.83 | 0.03 | 0.88 | 0.87 | -0.01 *** | 1.02 | 0.76 | -0.26 *** | 4.5% |
| Financial Specialists | 0.87 | 0.74 | -0.13 ** | 0.86 | 0.82 | -0.04 | 1.03 | 1.06 | 0.03 *** | 1.28 | 0.89 | -0.38 *** | 3.5% |
| Production | 1.43 | 1.43 | 0.00 * | 1.02 | 1.05 | 0.03 *** | 0.94 | 0.94 | 0.00 * | 1.05 | 0.91 | -0.14 *** | 2.8% |
| Technicians | 0.70 | 0.90 | 0.20 *** | 0.78 | 0.65 | -0.13 *** | 1.08 | 1.05 | -0.03 *** | 1.06 | 0.88 | -0.18 *** | 2.4% |
| Protective Service | 0.53 | 0.47 | -0.06 *** | 1.28 | 1.35 | 0.07 *** | 1.02 | 1.00 | -0.02 *** | 1.44 | 1.20 | -0.25 *** | 2.3% |
| Education, Training, and Library | 0.78 | 0.77 | -0.01 * | 1.28 | 1.18 | -0.10 ** | 1.01 | 1.03 | 0.02 * | 1.35 | 1.21 | -0.14 * | 1.8% |
| Arts, Design, Entertainment, Sports, and Media | 0.94 | 0.87 | -0.06 *** | 0.92 | 0.96 | 0.04 | 1.00 | 1.02 | 0.02 *** | 1.13 | 1.04 | -0.09 * | 1.4% |
| Farming, Fisheries, and Forestry | 2.94 | 2.70 | -0.25 *** | 0.26 | 0.31 | 0.06 *** | 0.78 | 0.83 | 0.05 *** | 0.83 | 0.81 | -0.02 * | 1.3% |
| Architecture and Engineering | 1.09 | 1.13 | 0.04 * | 0.58 | 0.51 | -0.07 ** | 1.02 | 1.02 | 0.00 *** | 1.07 | 0.84 | -0.23 *** | 1.2% |
| Legal | 0.30 | 0.28 | -0.02 * | 1.14 | 1.11 | -0.02 | 1.20 | 1.22 | 0.02 *** | 1.81 | 1.28 | -0.52 *** | 0.8% |
| Construction | 2.25 | 2.33 | 0.09 ** | 0.49 | 0.47 | -0.03 * | 0.87 | 0.87 | 0.00 *** | 1.06 | 0.96 | -0.10 *** | 0.8% |
| Extraction | 0.67 | 0.67 | 0.00 | 0.30 | 0.43 | 0.14 *** | 1.13 | 1.18 | 0.05 *** | 0.88 | 1.24 | 0.36 *** | 0.1% |

*=statistically significant change in denominator, **= statistically significant change in numerator, ***= statistically significant change in denominator and numerator

Appendix 2. Table 3. Occupational Crowding Index for Women using CPS Data, Before and During COVID

| Occupation Group | Immigrants | | | U.S. Born Workers | | | | | | | | | % in frontline industries |
|---|---|---|---|---|---|---|---|---|---|---|---|---|---|
| | | | | Black | | | White Non-Hispanic | | | Hispanic | | | |
| | Pre-COVID | During COVID | Difference | Pre-COVID | During COVID | Difference | Pre-COVID | During COVID | Difference | Pre-COVID | During COVID | Difference | |
| Healthcare Support | 1.39 | 1.40 | 0.01 * | 1.66 | 1.68 | 0.02 * | 0.80 | 0.78 | -0.02 *** | 1.25 | 0.77 | -0.49 *** | 67.8% |
| Healthcare Practitioners and Technicians | 0.83 | 0.79 | -0.04 *** | 1.06 | 1.00 | -0.06 *** | 1.03 | 1.05 | 0.01 * | 0.97 | 1.04 | 0.07 *** | 57.7% |
| Community and Social Services | 0.49 | 0.53 | 0.04 ** | 2.13 | 1.90 | -0.23 ** | 0.94 | 0.96 | 0.02 * | 1.67 | 0.61 | -1.06 * | 45.8% |
| Building and Grounds Cleaning and Maintenance | 3.35 | 3.30 | -0.05 * | 0.87 | 0.79 | -0.08 *** | 0.58 | 0.59 | 0.01 * | 0.85 | 1.24 | 0.39 * | 44.0% |
| Personal Care and Service | 1.65 | 1.62 | -0.03 * | 1.14 | 1.04 | -0.10 *** | 0.85 | 0.85 | 0.01 * | 1.02 | 0.93 | -0.10 *** | 40.0% |
| Transportation and Material Moving | 1.57 | 1.65 | 0.08 *** | 1.68 | 1.53 | -0.15 *** | 0.79 | 0.77 | -0.02 *** | 1.04 | 0.90 | -0.14 *** | 27.8% |
| Office and Administrative Support | 0.70 | 0.72 | 0.01 *** | 1.01 | 1.03 | 0.03 *** | 1.03 | 1.03 | -0.01 *** | 1.21 | 0.83 | -0.38 *** | 19.8% |
| Management in Business, Science, and Arts | 0.82 | 0.79 | -0.03 ** | 0.78 | 0.86 | 0.08 *** | 1.07 | 1.06 | 0.00 *** | 0.91 | 1.09 | 0.18 *** | 14.1% |
| Sales and Related | 0.95 | 0.95 | 0.00 * | 0.87 | 0.86 | -0.01 * | 1.02 | 1.02 | 0.01 * | 1.14 | 0.93 | -0.21 * | 12.6% |
| Life, Physical, and Social Science | 1.30 | 1.20 | -0.09 ** | 0.70 | 0.67 | -0.03 * | 0.95 | 0.97 | 0.02 * | 1.25 | 0.95 | -0.30 *** | 11.3% |
| Business Operations Specialists | 0.77 | 0.82 | 0.05 ** | 1.02 | 1.03 | 0.00 * | 1.02 | 1.02 | -0.01 *** | 1.06 | 0.95 | -0.11 *** | 9.6% |
| Food Preparation and Serving | 1.81 | 1.83 | 0.02 * | 1.11 | 1.07 | -0.04 * | 0.81 | 0.82 | 0.01 * | 1.06 | 0.99 | -0.07 * | 8.8% |
| Installation, Maintenance, and Repair | 0.87 | 0.94 | 0.07 * | 1.31 | 1.13 | -0.18 *** | 0.96 | 0.97 | 0.01 * | 1.08 | 0.85 | -0.23 * | 8.0% |
| Education, Training, and Library | 0.59 | 0.60 | 0.01 | 1.04 | 1.04 | 0.00 | 1.08 | 1.08 | 0.00 *** | 1.22 | 0.85 | -0.37 * | 7.7% |
| Computer and Mathematical | 1.63 | 1.78 | 0.16 ** | 1.26 | 1.17 | -0.09 ** | 0.82 | 0.80 | -0.02 *** | 0.88 | 1.17 | 0.29 * | 6.8% |
| Farming, Fisheries, and Forestry | 3.39 | 2.97 | -0.42 ** | 0.35 | 0.21 | -0.13 ** | 0.60 | 0.71 | 0.11 *** | 0.95 | 0.97 | 0.02 *** | 5.7% |
| Production | 2.28 | 2.25 | -0.03 * | 1.23 | 1.23 | 0.00 * | 0.73 | 0.73 | 0.00 * | 0.94 | 1.12 | 0.19 * | 5.5% |
| Financial Specialists | 0.94 | 1.03 | 0.08 *** | 1.01 | 1.08 | 0.07 *** | 1.00 | 0.96 | -0.04 *** | 0.96 | 0.93 | -0.03 *** | 5.5% |
| Protective Service | 0.51 | 0.46 | -0.04 * | 2.70 | 2.53 | -0.17 *** | 0.78 | 0.81 | 0.03 * | 1.40 | 0.71 | -0.69 * | 3.6% |
| Arts, Design, Entertainment, Sports, and Media | 0.80 | 0.83 | 0.03 * | 0.54 | 0.67 | 0.13 *** | 1.08 | 1.06 | -0.02 *** | 1.13 | 0.97 | -0.16 * | 3.3% |
| Technicians | 1.21 | 0.72 | -0.49 *** | 1.12 | 0.86 | -0.26 *** | 0.96 | 1.05 | 0.10 *** | 0.77 | 1.27 | 0.51 * | 3.2% |
| Legal | 0.60 | 0.59 | -0.01 * | 0.98 | 1.04 | 0.06 *** | 1.05 | 1.05 | 0.00 * | 1.33 | 0.78 | -0.55 * | 1.7% |
| Architecture and Engineering | 1.47 | 1.73 | 0.26 ** | 0.68 | 0.57 | -0.11 ** | 0.94 | 0.87 | -0.08 *** | 0.70 | 1.05 | 0.35 *** | 1.3% |
| Construction | 1.89 | 2.16 | 0.27 *** | 0.54 | 0.55 | 0.01 * | 0.89 | 0.84 | -0.06 *** | 0.98 | 1.00 | 0.02 * | 1.0% |
| Extraction | 0.68 | 1.09 | 0.41 * | 0.75 | 0.43 | -0.32 * | 1.20 | 1.23 | 0.03 * | 0.00 | 5.26 | 5.26 * | 0.0% |

*=statistically significant change in denominator, **= statistically significant change in numerator, ***= statistically significant change in denominator and numerator